# Regulating for AI Legitimacy

**Gilad Abiri**

## I. Introduction

AI systems already govern. They rank speech and allocate attention; they filter applicants and triage claims; they recommend, flag, and withhold. This is likely only the prelude (Lazar, 2024). Technical improvements—what the field calls “alignment” —do not decide whether people will accept that governance. On the sociological view used here, legitimacy turns on recognition: whether power appears rule-bound, justified, and situated in institutions that speak for those it governs (Beetham, 1991; Weber, 1968). Recent experience with social media and search—forms of algorithmic power long before generative models—shows the stakes. Distribution got faster and broader even as publics questioned authorization and reasons, producing recurrent social and political conflicts grounded in the question of who gets to set the rules for speech and attention (Abiri & Guidi, 2023; Cohen, 2017; Klonick, 2018; Suzor, 2019). As general-purpose AI embeds upstream of many domains, the risk is to repeat that drama at larger scale: systems that perform yet lack legitimacy, inviting resistance, regulatory backlash, and governance paralysis.

Standard accounts of AI governance miss the legitimacy question in two ways. First, they often collapse authority into performance, treating safety audits, benchmarks, or risk scores as a clear indication of acceptability (Stone & Mittelstadt, 2024). But publics do not simply ask whether outputs are accurate or efficient; they ask whether the power to decide is exercised rightfully—according to rules they recognize, for reasons they can follow, with avenues to challenge and repair (Beetham, 1991; Tyler, 2006). Second, where governance is framed as rights compliance or product regulation, the sociological dimension of legitimacy—audience beliefs formed through visible authorship, intelligible reasons, and credible second looks—remains under-specified (Jasanoff, 2005; Post & Siegel, 2004). The result is a gap between technically careful systems and publics who resist, doubt, or disengage. The point is not that alignment and legitimacy are unrelated, but that it's perfectly possible to live in a world where a perfectly benevolent AI meets the political maelstrom of a full-blown legitimacy crisis. Performance doesn't guarantee recognition. This matters normatively and practically: voluntary compliance, cooperation, and durable policy all depend on recognition, not just performance (Tyler, 2006; Post, 2012).

The Article's contribution is to make legitimacy an explicit regulatory aim for AI. The claim is modest: law cannot manufacture trust or substitute for safety, but it can structure recognition—turning performance into reasons, and reasons into authority that publics can potentially accept. Building on central legitimacy studies in political sociology, science-and-technology studies, and public-law theory, I distill three audience-facing principles for stucturing regulating for AI Legitimacy. The principle of Integration locates AI rule-setting in venues a polity already treats as authoritative, so authorship is visible and situated (Post, 2012; Post & Siegel, 2004; Habermas, 1996). The Principle of Familiarity presents rules and reasons in locally credible forms—what Jasanoff calls civic epistemologies—so people can recognize, anticipate, and navigate how they are governed (Dewey, 1927; Jasanoff, 2005). The Principle of Contestation guarantees a credible second look through accessible challenges, impartial review, and effective redress, aligning institutional design with what procedural-justice research shows about acceptance (Tyler, 2006; Schauer, 1995; Mashaw, 2018). These principles translate and refine the caution from platform governance: absent recognized authorship, audience-understandable reasons, and real review, algorithmic rule is experienced as arbitrary—even when it "works" (Abiri & Guidi, 2023; Cohen, 2017).

The Article proceeds as follows: Part II clarifies the sense of legitimacy at work—sociological recognition, not a technical property—and distinguishes it from alignment, using public-facing algorithmic systems to show why both matter. Part III maps where AI legitimacy falters today: opacity that blocks audiences from forming justified beliefs; the political-integration problem, where private actors exercise public-facing authority without recognizable authorization; and administrative strains when automation is folded into the state. Part IV considers how legality can be used to repair these deficits: thin/formal legality supplies visible forms that signal non-arbitrariness—publication, stability, consistent application—while thick/constitutional legality adds public authorship, audience-facing reasons, and contestability, with scope conditions and limits. Part V distills the analysis into three principles for practice—integration, familiarity, and contestation—offered as portable guidance different jurisdictions and sectors can adapt, not as a universal fix.

The payoff is twofold. For policy, the principles suggest a layered program jurisdictions can tailor to local circumstances: seat consequential AI rule-sets in recognized venues of authorship and revision; build public-facing rulebooks, stable change logs, and statements of reasons in locally credible forms; and guarantee an impartial second look with published timelines, access for reviewers, and remedies that fix both outcomes and rules. These are not merely compliance tasks; they are the conditions under which publics can see rules, hear reasons, own authorship, and use avenues of contestation. For theory, the Article reframes AI governance as a problem of rightful authority, not only of risk or rights. If legitimacy is recognition in context, then law’s distinctive contribution is to provide the forums, genres, and procedures through which performance becomes justification and justification becomes authority. That is a modest claim, but it is actionable—and it is how AI can be governed in ways that publics recognize as rightful.

## II. Why Legitimacy (not just "alignment")?

Artificial intelligence already allocates attention, opportunities, and burdens across social life. It amplifies or suppresses speech on social platforms, structures the informational field through ranking and retrieval in search, and informs resource allocation and administrative decisions across public and private institutions. In short, AI does not merely assist; it governs (Lazar, 2024). Following David Beetham, I use *legitimacy* in a sociological sense: the widespread belief among those subject to power that it is exercised rightfully or justifiably (Abiri, 2025, p. 605; Beetham, 1991). Legitimacy, in this essay, is therefore an audience-side phenomenon—about recognition, not metaphysical correctness.

Distinguishing legitimacy from alignment follows. Alignment asks whether systems optimize for the "right" objectives, reduce risk, and behave safely under evaluation. Legitimacy asks by what right those objectives are set and enforced, and whether those who live under the system accept that authority as justified. Conflating the two—treating a well-aligned system as ipso facto legitimate—mistakes performance for authorization (Stone & Mittelstadt, 2024). This distinction is well supported across adjacent traditions: Fallon (2005) characterizes constitutional legitimacy as an active belief that claims to authority deserve respect or obedience; Tyler (2006) shows that people comply because they regard authorities and procedures as proper, not only because outcomes benefit them; Weber (1968) describes modern legal-rational rule as depending on people's beliefs about the propriety of rule-bound institutions. Whatever their differences, these accounts converge on the sociological core: legitimacy is about whether a governed public recognizes power as justified (Fallon, 2005; Tyler, 2006; Weber, 1968).

We already know, moreover, that alignment and legitimacy come apart—and that both matter—because we have lived through it with social media. Platforms undeniably delivered a net improvement on familiar media metrics: vastly expanded access, lower barriers to speak and organize, unprecedented distribution capacity, and a flourishing of user creativity. Yet those gains were paralleled by a trust and legitimacy crisis, as publics questioned why a handful of firms should rule the conditions of online visibility and speech. In work I Guidied, we described this as a legitimation crisis with tangible consequences: platforms faced real prospects of audience withdrawal and intensifying regulation (Abiri & Guidi, 2023, pp. 92–96). Media-law scholarship has likewise recast platform rule as a species of *governance*, not mere product management (Klonick, 2018; Suzor, 2019). The lesson is not that functionality or “alignment” is unimportant, but that usefulness ≠ authorization. A system can improve outcomes yet still be illegitimate if those subject to it do not regard its rule as rightful (see also Abiri, 2025, p. 605).

Social media also shows that the algorithmic legitimacy problem is not confined to government. These are private institutions exercising power in ways the public experiences as quasi-public—most starkly, content moderation over what often feels like *all of human speech*. In our analysis, we emphasized why private claims to govern the public sphere ring hollow: platforms are “too public to be fully private and too concerned with profit to be believed to act in the public interest” (Abiri & Guidi, 2023, p. 94). Legal scholars of the platform economy long ago noted that the largest intermediaries exercise infrastructural, public-facing authority (Cohen, 2017). And the same drama plays out in search. Search engines are not mere conduits; they are also algorithmic power. They shape what we may plausibly come to know and when, determining access to public-relevant information about matters such as voting and public health (Abiri, 2025, pp. 601–603). As Seth Lazar puts it, these are “automatic authorities”—automated systems that exercise power by substantially determining *what we may know, what we may have, and what our options will be* (Lazar, 2024). Seen this way, social media and search are not peripheral to “AI governance”; they are AI governance in everyday life—central instances of algorithmic authority whose legitimacy must be assessed in the sociological sense at stake here.

Taking stock, the platform era offers a working proof of concept: alignment gains without legitimacy invite political and social blowback; legitimacy cannot be presumed from utility. Private actors can—and routinely do—exercise public-facing algorithmic power that calls for public recognition as justified authority, even outside formal government. Public-law scholars have made closely related points in the administrative context: when agencies automate, legitimacy problems arise that cannot be fixed by accuracy alone (Calo & Citron, 2021; Engstrom et al., 2020). That is why, throughout this Article, I treat legitimacy as an autonomous regulatory objective alongside alignment. The question is not only *whether* systems optimize desirable goals, but *by what right* they govern the informational and social environments in which we live (Abiri & Guidi, 2023, pp. 92–96; Abiri, 2025, pp. 601–606; Cohen, 2017; Klonick, 2018; Lazar, 2024). The institutional pathways for securing that recognition are taken up in the sections that follow.

## III. The Legitimacy Challenges of AI Power

This section examines where AI authority faces legitimacy problems that go beyond simple failures of performance. The sociological framework used here treats recognition as dependent on whether governance appears reasoned, authored, and answerable. Three persistent problems organize the analysis. First, opacity: when systems cannot provide comprehensible justifications—even when their decisions are correct—those determinations look arbitrary, eroding the foundational beliefs that legitimacy requires. Second, political integration: private entities now set norms governing speech, visibility, and access at massive scale, often outside the institutional channels that the public recognizes as conferring authority. This creates power without recognized authorship. Third, administrative automation: when agencies deploy AI in ways that weaken reason-giving, participation, documentation, and appellate review, they risk more than errors—they undermine the state's right to govern through law. The subsections that follow address these as connected dimensions of one central challenge: without justifications, identifiable authorship, and meaningful reconsideration, algorithmic governance becomes governance by no one.

**A. Opacity**

Legitimacy, in the sense I use here, turns on whether those subject to power can recognize it as rightfully exercised. That recognition depends, in part, on receiving reasons they can understand and evaluate. Procedural-justice research shows that people's willingness to accept decisions—even adverse ones—rises when authorities provide intelligible explanations that signal neutrality, voice, and respect (Tyler, 2006). If those who are governed cannot tell why a system acted, even accurate outcomes read as arbitrary—undermining the audience beliefs on which legitimacy rests.

Modern machine learning compounds this problem. AI opacity has at least two distinct sources. First, complexity opacity: non-intuitive, high-dimensional statistical inference that resists lay (and often expert) understanding. Second, proprietary opacity: the deliberate withholding of model internals, data, and prompts under trade-secret and IP doctrines (Chesterman, 2021). These are different in law but similar in effect: both frustrate publicity and reason-giving, making it harder for affected audiences to see decisions as grounded in acceptable reasons rather than in inscrutable machinery.

Democratic theory makes the same point in institutional terms. When public authorities—or private actors exercising public-facing power—cannot connect an outcome to relevant facts and applicable rules in a way that is accessible to those affected, the decision fails a basic test of democratic legitimacy (Beckman, Hultin Rosenberg, & Jebari, 2024). The problem is not only whether a reason exists somewhere inside a pipeline, but whether the audience can grasp it well enough to judge that the exercise of power is justified.

A parallel argument comes from recent work on “automatic authorities.” If algorithmic systems already shape what we may know, what we may have, and what our options will be, then those who design and deploy them incur duties of explanation: they owe their publics reasons sufficient for a reasonable person to determine that they are governed legitimately (Lazar, 2022). This reframes explainability from a usability preference into a legitimacy requirement: explanations are one of the few bridges between technical performance and audience recognition.

I have elsewhere made this point expressly in the AI context: opacity is a legitimacy deficit. When the governing logic of an AI system is not visible in audience-understandable terms, publics cannot form the beliefs that confer rightful authority—regardless of the system’s measured accuracy (Abiri, 2025). Treating opacity as a legitimacy defect, rather than merely an engineering quirk, properly reorients regulatory priorities toward the conditions under which audiences can recognize AI-backed authority as justified.

Meeting that requirement does not mean every single output must be mechanically dissectible. What matters for legitimacy is that the governing framework be public-facing and reason-giving in ways ordinary audiences can understand. Where per-decision explanations are infeasible, credible substitutes are available: publish stable, system-level principles that state what the model is trying to do and what it will not do; provide short, plain-language reason letters when decisions carry significant consequences; create predictable second-look channels that let people contest outcomes and receive responsive reasons. Each of these speaks directly to the recognition mechanisms identified by Tyler (2006): they preserve voice, signal neutrality, and—above all—supply explanations that help recipients judge decisions as properly grounded.

The upshot is straightforward. Opacity is a legitimacy defect. If audiences cannot access reasons—or credible substitutes for them—they cannot form the beliefs that confer rightful authority. The regulatory and institutional program that follows therefore treats audience-understandable reasons as a first-order requirement of algorithmic governance, alongside accuracy and safety (Abiri, 2025; Beckman et al., 2024; Chesterman, 2021; Lazar, 2022; Tyler, 2006).

**B. Private AI Power**

Much of the algorithmic power that structures contemporary life is exercised by private institutions whose rule-setting looks and feels public in its effects. Platforms decide whose speech is amplified or restricted and set the terms of visibility in the digital public sphere; search and recommender systems shape what people may plausibly come to know; and frontier model providers increasingly pre-commit the “constitutions,” guardrails, and safety layers that condition downstream speech and access at global scale (Abiri, 2025, pp. 601–603, 648–652). The upshot is de facto public authority without the ordinary pathways of public authorization.

This authority is classically governance rather than mere product design. Private intermediaries set and enforce rules of speech, visibility, association, and opportunity—public-facing functions by any other name (Cohen, 2017; Klonick, 2018; Suzor, 2019). Content policies and ranking frameworks operate as institutionalized rule systems that allocate attention and constrain behavior at scale; they are experienced by users as legal-like in aspiration and effect (Abiri & Guidi, 2023, pp. 110–116). When private actors perform these public functions, audiences naturally evaluate them by public standards of rightful rule—and notice when authorship by a political community is missing.

Private legitimation strategies have tried to meet that expectation by mimicking familiar public institutions. Platforms formalize internal hierarchies, write rules in quasi-legal idioms, professionalize decision-makers, and offer appeals—sometimes with external review by a court-like body (Abiri & Guidi, 2023, pp. 112–116). Meta's Oversight Board is the emblematic case: a chartered, quasi-judicial layer that issues reasoned opinions and aspires to precedent (Oversight Board, 2019). These experiments show promise—procedures can reduce arbitrariness, discipline internal reasoning, and signal neutrality (Klonick, 2018). They also expose limits: legal mimicry can legitimate how content moderation is exercised, not why platforms—or their boards and CEOs—should hold unparalleled power over the digital public sphere (Abiri & Guidi, 2023, pp. 122–135; Douek, 2024). Put differently, procedural polish cannot substitute for public authorship.

The constraint is general. Rory Van Loo argues that even as corporations adopt procedural-justice features to enhance fairness and accountability, they cannot escape the fundamental fact that their authority is not derived from the consent of the governed (Van Loo, 2016, pp. 560–562). In the AI setting, the point travels: private constitutionalization of model behavior may improve procedure yet still lack a credible link to an identifiable political community; absent that link, claims to rightful authority remain fragile (Abiri, 2025, pp. 640–652).

The regulatory moral follows. Platform and model governance should be treated as sites of public rule-setting that demand public authorship—or credible substitutes for it. Corporate proceduralism can mitigate arbitrariness and teach organizations to reason publicly, but it cannot, by itself, supply authorization. Closing the political-community gap therefore requires integration of private AI power into frameworks that can plausibly claim public authorship: co-regulatory architectures, binding public-law constraints on platform and model rulebooks, and channels through which affected publics can contest and shape the governing rules (Cohen, 2017; Finck, 2018; Suzor, 2019).

### C. Public-Sector Automation and Administrative Legitimacy

Inside the state, legitimacy is not a luxury add-on to accurate decision systems; it is tied to administrative law—the background principles that make coercive governance recognizably lawful (legality, reasoned decisionmaking, transparency, participation, and review). Sunstein and Vermeule describe how modern administrative law embodies these rule-of-law commitments across doctrines, from notice-and-comment and reason-giving to non-arbitrariness and reviewability. Any turn to automation that dulls these commitments risks not merely error but a deficit in the authority of the state's actions as such (Sunstein & Vermeule, 2018).

Recent experience shows how automation can strain those commitments. Calo and Citron catalog how agency uses of machine learning and automated tools can short-circuit due process (opaque eligibility determinations and sanctions), reason-giving (outputs without comprehensible explanations), participation (rulemaking by model iteration rather than public comment), and review (thin or non-existent records capable of judicial scrutiny). Their bottom line is expressly about legitimacy: absent procedural repair, automated administration risks eroding the very grounds on which citizens accept agency authority (Calo & Citron, 2021).

This is not an abstract worry. The Administrative Conference's Government by Algorithm report canvassed AI use across 142 federal departments, agencies, and sub-agencies, then drilled into case studies at entities like the SEC, SSA, USPTO, FDA, FCC,

CFPB, and USPS. The authors identify cross-cutting institutional and legal challenges—documentation, transparency, explainability, procurement, and oversight—that recur across domains. The scale and diversity of use confirm that legitimacy questions are already a systemic feature of public administration, not a future hypothetical (Engstrom, Ho, Sharkey, & Cuéllar, 2020).

The administrative-law literature points to concrete levers for procedural repair. Citron (2008) argued early that automation upends traditional procedural safeguards and requires a technological due process attuned to code-driven adjudication and rulemaking. Coglianese and Lehr (2017) examine how machine-learning decision tools intersect with core administrative and constitutional doctrines, outlining conditions under which "regulating by robot" can remain compatible with reason-giving and review. Courts, too, have a role: Deeks (2019) contends judges should sometimes demand explanations for algorithmic decisions to make judicial review meaningful. And the lived consequences of administrative automation—documented in Eubanks (2018)—underscore why legitimacy cannot be secured by accuracy alone; opacity and thin process trigger resistance, error cascades, and loss of trust (Citron, 2008; Coglianese & Lehr, 2017; Deeks, 2019; Eubanks, 2018).

Regulators have begun to encode these legitimacy conditions. OMB Memorandum M-24-10 now requires agencies to implement risk-management practices, public consultation, notice, monitoring, human consideration and remedies for adverse decisions, and other safeguards for rights-impacting AI uses—explicitly tying agency AI to transparency, oversight, and public participation. ACUS Statement #20 (Agency Use of AI) likewise instructs agencies to address transparency, harmful bias, decisional authority, and oversight when adopting or modifying AI systems (OMB, 2024; ACUS, 2021; see also 2024–2025 agency compliance plans).

For a legitimacy-first approach within agencies, accuracy is necessary but insufficient. What must be shored up are the procedures that make administrative power recognizably rightful: intelligible reason-giving calibrated to audiences (so recipients can understand the "why"), records robust enough for arbitrary-and-capricious review, opportunities for voice in rule formation (modernized to fit data-driven systems), and oversight practices that keep human accountability legible even when models scale. This is precisely the work of aligning automation with the morality of administrative law—not by resisting tools per se, but by insisting that their use preserve publicity, reasoned justification, and reviewability (Sunstein & Vermeule, 2018; Calo & Citron, 2021; Engstrom et al., 2020).

Put simply: administrative legitimacy is the state's durable claim to govern through rules and reasons. Automation that sidelines reasons, bypasses publics, or frustrates review does not just risk mistakes; it threatens that claim. The regulatory program that follows therefore treats procedural repair—reason-giving, participation, and review suited to automated pipelines—as a first-order requirement of public-sector AI, alongside accuracy and efficiency (Engstrom et al., 2020; Calo & Citron, 2021).

## IV. Legal Legitimation Strategies for AI

Parts I and II identified the problem space: AI systems already govern—they set and enforce rules over speech, attention, access, and opportunity—and publics evaluate that governance in terms of legitimacy. Part III asks the practical follow-up: what can legal legitimation tools contribute to regulating for AI legitimacy?

I distinguish between thin and thick legality. I distinguish between thin and thick legality. By *thin legality* I mean the forms of law: public and stable rules, prospectivity, consistent application, and professionalized routines—the recognizable juridical signals of order and non-arbitrariness. These forms can, in some contexts, raise perceived legitimacy by making governance legible and serious, even before deeper questions of authorization are settled (Fu, Xu, & Zhang, 2021; Weber, 1968; Zhang, 2024).

By *thick legality* I mean the substance of law as publicly authored and owned—rules that are justified to, and issued from within, a political community, coupled with audience-facing reason-giving and opportunities for voice and contestation. On the sociological account here, thick legality anchors legitimacy because it links rule to recognized authorship and acceptable reasons (Abiri, 2025; Beetham, 1991; Tyler, 2006).

The remainder of Part III proceeds in two steps. Section III.A evaluates what thin/formalist legality can and cannot accomplish for sociological legitimacy—its scope conditions, limits, and the risks of legitimacy-washing if form substitutes for substance. Section III.B turns to constitutional/participatory strategies, showing how public authorship, explanation, and contestation address the legitimacy challenges surfaced in Part II.

**A. "Thin" Legality**

This subsection asks whether thin legality—the *forms* of law (publicly articulated and stable rules, prospectivity, consistent application, professionalized decision routines)—can bolster *sociological* legitimacy for AI even before we secure a thicker settlement grounded in public authorship. The claim is modest but useful: visible legality can move audience beliefs about the rightfulness of power by signaling non-arbitrariness and institutional seriousness. It is not sufficient, but it is often instrumental for a regulatory program that aims to make AI governance recognizably rightful.

Authoritarian and quasi-authoritarian contexts are helpful laboratories for isolating form's legitimation effect—not because they are normative models, but because thin legality can be observed largely apart from thick, community-grounded authorization. In those settings, the signal value of legality is unusually clear. Zhang argues that orderly legal form— publication, consistency, professionalized decision-making—can operate as an independent source of sociopolitical legitimacy (Zhang, 2024). Fu, Xu, and Zhang supply direct evidence: "pure" legality (stripped of rights guarantees and independent courts) raises perceived legitimacy among respondents in China, even for controversial policies such as online speech censorship; at the same time, its effect is smaller than audience-facing fairness cues associated with procedural justice (voice, respectful treatment, neutrality) (Fu, Xu, & Zhang, 2021; Tyler, 2006).

Why does thin legality move belief? Part of the answer is signaling. In Weber's account of legal-rational authority, publics credit governance that presents itself as *rule-bound* and *office-based*, reading those cues as evidence of non-arbitrariness (Weber, 1968). Fuller's "inner morality of law"—generality, publicity, prospectivity, clarity, relative stability—explains why these formal features read as governance with reasons, even before deeper questions of authorship are resolved (Fuller, 1969). Organizational sociology adds a complementary mechanism: organizations adopt formal structures as legitimacy devices—myth and ceremony that reassure key audiences (Meyer & Rowan, 1977)—and they converge through isomorphism on recognizable, professionalized templates that broadcast competence and propriety (DiMaggio & Powell, 1983). In short, thin legality looks like rightful rule: there are rules; the rules are public; officials follow them.

Concretely for AI, thin legality mitigates several endemic problems. Against opacity, legal form pushes toward documentation and publication: model/system cards, datasheets for datasets, and standardized disclosures make governing assumptions and limits visible and more contestable in public fora (Mitchell et al., 2019; Gebru et al., 2021). Against drift and rapid iteration, legal form favors versioning, change-logs, and release discipline, anchoring expectations and enabling after-the-fact review (NIST AI RMF 1.0; Generative AI Profile). Against fragmented practices, formal classification regimes (e.g., risk-tiering) can align incentives and make compliance observable across actors (EU AI Act; Canada's Algorithmic Impact Assessment under the Directive on Automated Decision-Making). And against procurement opacity, legal form can require minimum documentation, audit trails, and professionalized roles as conditions of deployment. None of these solve the inherent opacity of ML models, but they make it so that, at scale, AI governance looks legible and rule-bound rather than ad hoc.

Putting AI under law's forms—codifying baseline obligations in statutes and regulations; publishing stable, versioned rule-sets for model behavior; professionalizing change-management, exception-handling, and oversight; keeping public registers and compliance files; documenting reasons in a standardized way—will likely yield a legitimation dividend by making governance appear legible and predictable. Members of the public tend to credit order and follow-through: when they can *see* the rules and observe decision-makers conforming to them, they are more inclined to view the exercise of power as justified. That said, performance still matters. Thin legality does not immunize bad policy, and legitimacy built on form alone is fragile—especially in democracies where audiences expect authorship, explanation, and avenues for contestation (Tyler, 2006).

There are also risks. Legal form can become decoupled from practice—what Edelman calls legal endogeneity—where organizations win deference by showcasing policies, trainings, and review boards that change little on the ground (Edelman, 2016). In AI governance, that dynamic looks like legitimacy-washing: publication without constraint, professionalization without accountability, ceremonies of consultation without voice. Thin legality raises perceived legitimacy when it credibly signals non-arbitrariness and adherence; it diminishes it when audiences read the signals as theater. This is a reason to pair formalism with basic quality controls and audience-facing explanation, so that form and substance reinforce one another rather than drift apart.

The upshot is a division of labor that sets up the next subsection. Thin legality (the forms of law) is necessary and often effective at quickly improving recognition: it makes AI appear legible, predictable, and professionally administered, and it typically yields a baseline bump in acceptance when AI is “put into law.” Thick legality (the substance of law) supplies what form cannot—public authorship/ownership, audience-understandable reason-giving, and contestability that tie AI rule to a political community. Part III.B turns to those constitutional/participatory tools; taken together, the two strands seek AI that is not only safer, but recognizably rightful.

## B. "Thick" Legality

Law’s distinctive claim to legitimacy is that it is authored by the political community. People accept legal rule not because it is orderly, but because they can see it as theirs—made and revised by institutions that speak for them (Beetham, 1991; Post, 2012; Post & Siegel, 2004; Habermas, 1996). Thick legality seeks to harness that source of recognition for AI by relocating consequential rulemaking to publicly authoring venues, so that claims of authority travel with the forums that issue and revise the rules rather than with the firms that implement them. There are multiple ways to do this. One example is **public constitutional AI**, which builds on Anthropic’s “constitutional AI” insight—steering via an articulated rule-set—while relocating authorship from company to polity (Abiri, 2025; Bai et al., 2022).

By public constitutional AI I mean a governance arrangement in which a model or model family is bound by a rule-set adopted in public, published with its reasons, and renewed or amended through public procedures (Abiri, 2025). The constitution states purposes and exclusion zones, articulates the value trade-offs that guide behavior, sets guardrails and criteria for exceptions, and specifies who may deviate, how, and under what justificatory burdens. Issuance and revision occur in venues the polity already recognizes as authorizing, so authorship—rather than internal compliance documentation—is visible and attributable. This is one design family within the broader project of thick legality; parallel approaches can use the same authorship logic without the “constitutional” label.

Venues for public authorship should track where the polity already locates law-making authority. For public deployments, authorship can proceed through statutes or agency rulemaking, with the constitutional text functioning as the publicly adopted rule-set. For private actors with public effects (platforms, search, frontier model providers), authorship can be anchored in regulator-supervised industry codes, municipal bodies, or recognized professional orders that carry public duties. For base-model providers with global reach, an upstream rule-set can be authored in national or municipal forums and then layered with local overlays, acknowledging that a single global settlement rarely attracts recognition across polities (Abiri, 2025).

Authorship must be credible rather than ceremonial. Deliberative-democratic designs supply feasible machinery—representative mini-publics, balanced adversarial briefings, moderated deliberation, and transparent decision rules—situated within a broader ecosystem of legislatures, agencies, courts, professional orders, and civil society (Fishkin, 2009, 2018; Landemore, 2020; Mansbridge et al., 2012; Fung, 2004). Scope conditions should be explicit: public authorship contributes to recognition where representation is credible, materials and proceedings are public and reviewable, and adopting institutions have a duty to consider, respond, and either adopt or justify deviations. Absent those conditions, constitutions (or any authorship device) read as ceremony.

Materiality thresholds help determine when to require public authorship. Constitutional rule-setting is warranted for systems that meaningfully affect speech and visibility, access and opportunity, or legal and administrative outcomes; lighter-touch publication and registration may suffice elsewhere. This keeps the mechanism proportionate while reserving the strongest authorship signal for domains where legitimacy most needs shoring up.

The term constitutional signals authorship, durability, and constraint: a commitment adopted and renewed in public; versioned and logged; and auditable against the public record. Methodologically, if constitutions steer models through prompts and guardrails, publicly authored rule-sets can be translated into those prompts and audited against them—making the lineage from public authorship to model behavior inspectable (Bai et al., 2022; Author, 2025).

Public authorship addresses Private AI Power directly. Today, platforms, search, and model providers often set public-facing rules from corporate venues that publics do not recognize as authorizing. Reseating those rules in forums a polity already trusts reattaches authority claims to the community that must recognize them (Post, 2012; Post & Siegel, 2004; Habermas, 1996; Beetham, 1991). Complementary elements—audience-facing reasons and a credible second look—secure intelligibility and answerability, so that performance is converted into justification and justification into authority.

## V. Principles for AI Legitimacy

Parts I and II explained why alignment is not enough: people judge AI-backed power by whether it is exercised rightfully, and legitimacy falters where reasons are opaque, where private actors wield public-facing authority without authorization, and where automation strains notice, participation, and review. Part III examined how legality can be used in this setting: thin legality supplies the visible forms that signal non-arbitrariness—publication, stability, consistent application, professionalized routines—while thick legality adds the substance that ties rule to a political community—public authorship.

This Part shifts from those institutional uses to general principles. Legal legitimacy is produced within particular polities, through their own institutions and ways of making reasons public. There is no single global design that can deliver recognition everywhere; principles can travel where forms cannot. What follows distills the analysis into three guides that jurisdictions and sectors can adapt: political integration, which anchors AI rule-setting in venues and standards a polity already recognizes as authoritative; familiarity, which communicates rules and reasons in locally credible ways so people can recognize, anticipate, and navigate how they are governed; and contestation and remedy, which preserves a credible second look through accessible challenges, impartial review, and effective redress.

**A. Integration**

The principle of integration is simple: anchor AI rule-setting in venues and standards a polity already recognizes as authoritative, so authorship is visible and located. On the sociological view of legitimacy used here, people treat power as rightful when rules are made and revised in places they understand to speak for them—not merely when systems perform well. Integration, then, is about where AI rules live and who is seen to author them. It operationalizes Beetham's strands—rule-conformity, justifiability, and consent—for non-state and hybrid settings (Beetham, 1991).

Robert Post's account of democratic legitimation holds that authority sticks when expert governance is situated in the institutions through which publics form and express political judgment; law claims allegiance when it is authored and revised in forums that people already take to speak for them (Post, 2012; Post & Siegel, 2004). Habermas adds that law's procedures of will-formation and review are the medium that connects administrative power to public justification, but always within a particular constitutional order (Habermas, 1996). Together, these accounts support local integration: place AI rule-setting inside the sites a polity already credits with making and revising rules, and claims of authority will travel with that placement.

Applied to AI, integration asks for a few spare moves. First, seat consequential rule-sets where authority already resides. In public deployments, that means statutes or agency rulemaking. For private actors with public effects, use regulator-supervised codes, municipal bodies, or recognized professional orders that already carry public duties. The point is venue recognition: rules should be adopted where people already expect law-like authorship and revision.

Second, make authorship explicit. Adopt and renew rule-sets through public procedures that connect them to a political community: structured input at adoption; statements of reasons that explain value trade-offs in accessible terms; scheduled amendment cycles; and a petition-for-change track with a duty to respond. These are familiar pathways of public authorship, continuous with popular constitutionalism (Kramer, 2004) and with the public constitutional AI approach developed earlier (Author, 2025).

**B. Familiarity**

Familiarity names a practical requirement of legitimacy: people need to be able to recognize, anticipate, and navigate how they are governed. It is not mere transparency. On the view taken here, publics judge rule as rightful when rules and reasons are made public in ways that fit their polity's trusted modes of public reasoning—its civic epistemologies—so that ordinary people can form stable expectations and act.

Sheila Jasanoff's account of civic epistemologies holds that legitimacy turns on how a polity publicly tests, validates, and renders reasons credible—through its own trusted genres (reasoned decisions, advisory opinions, white papers), gatekeeping institutions (agencies, courts, academies, professional orders), and rituals of disclosure and critique (hearings, comment-and-response, expert testimony) (Jasanoff, 2005). This is related to how Dewey sees publics as acquiring practical bearings through repeated contact with stable signposts—rules, categories, and language that make governance navigable rather than opaque—so explanations must be cast in forms that people can learn and reuse over time (Dewey, 1927). Tyler supplies the uptake mechanism: when audiences receive clear, audience-understandable explanations delivered with cues of neutrality and respect, willingness to accept governance increases even where outcomes are not preferred (Tyler, 2006). Together, these accounts support the familiarity principle: present AI rules and reasons in a polity's own civic epistemology—plain-language rulebooks, stable versioning and change logs, statements of reasons keyed to recognized standards, and recurring public forums that teach and test those reasons—and recognition is far more likely to form than with disclosure alone.

Translated to AI, familiarity asks for a public-facing information architecture that people can actually learn. For consequential systems, publish rulebooks in plain language: the purposes of the tool, the core principles that guide decisions, exclusion zones, and the factors that predictably matter. Keep the vocabulary consistent and the categories stable enough for users, researchers, and officials to internalize. Pair the rulebook with versioned change logs and advance notice for material shifts, so people can see what changed, why, and when new expectations take effect. When major updates alter trade-offs, issue a short statement of reasons that connects the change to publicly articulated purposes and constraints—an explanation keyed to the polity's familiar genres.

Familiarity also requires layering. Not everyone needs the same level of detail, but everyone needs a credible starting point. Provide concise summaries for general publics, and link to deeper materials for advocates, watchdogs, and researchers. Use a consistent location and format so people know where to look—one public register, one canonical URL, one predictable document structure. Stability here is part of the signal: if the signposts move or the terms keep changing, people cannot learn the governance they are supposed to live under.

There is a pedagogical dimension. Familiarity grows when institutions take responsibility for explanation. Schools, public bodies, companies, and civil intermediaries should offer recurring, non-promotional primers on how the tools work in broad terms, where they are used, what values they embed, and how they are governed. Deliberative programs can deepen this: when people have time, materials, and a structured opportunity to learn, their judgments become more considered and their acceptance of hard trade-offs can increase (Fishkin, 2018). The aim is not universal technical literacy; it is a shared civic vocabulary in which reasons can be given and contested.

The design details matter. Write for the likely reader: short sentences, defined terms, concrete examples. Keep core categories stable; when change is necessary, publish migration guides that map old to new. Indicate how public input shaped outcomes—what was proposed, what changed, and why. Report at intervals on how rules performed against their stated aims, closing the loop between publication and practice. These follow-through cues convert disclosure into familiarity and make it easier for audiences to form the stable expectations that recognition requires.

Experience also counsels caution. Formal virtues—publicity, clarity, stability—help, but they are not self-executing. Transparency without structure can confuse or alienate; disclosure that exposes dysfunction without a path to improvement can depress trust; documents that drift from practice read as theater (Fuller, 1969; Scott, 1998; Grimmelikhuijsen & Meijer, 2014; Bauhr & Grimes, 2014). The remedy is craft, not volume: fewer documents that the public can actually use; consistent formats; real explanations tied to recognizable standards; and proof of learning over time.

Familiarity is modest in ambition and demanding in execution. It assumes that thin legality's forms—publicity, clarity, stability—are necessary conditions for recognition, then insists that those forms be occupied by reasons that fit local epistemic habits. The practical test is straightforward: can ordinary people find the rules, grasp what the tool is doing in their world, and foresee outcomes well enough to act? If the answer is yes, familiarity is doing its work.

**C. Contestation**

Contestation names the lived experience of rightful rule: a credible second look. People accept decisions when they can question them, obtain an impartial reconsideration on defined timelines, and receive an explanation and remedy that make sense in publicly declared terms. The aim is not to guarantee agreement but to make authority answerable in ways audiences recognize.

Tom R. Tyler's account of procedural justice holds that acceptance turns on whether people experience voice, neutrality, respectful treatment, and intelligible explanations (Tyler, 2006). Frederick Schauer adds that reason-giving disciplines officials and externalizes justification, making review possible and evaluable (Schauer, 1995). Jerry Mashaw treats reasoned administration as a core of democratic legitimacy, while Sunstein and Vermeule frame the morality of administrative law—reasons, participation, and review—as the medium that connects power to public justification in concrete forums (Mashaw, 2018; Sunstein & Vermeule, 2018). Philip Pettit and Pierre Rosanvallon supply the republican mechanism: institutions of contestation, oversight, and rectification keep authority continuously answerable, not merely accurate (Pettit, 1997/2012; Rosanvallon, 2008). Together, these accounts support contestation: provide clear channels to challenge, a genuinely impartial second look on defined timelines, and remedies tied to public reasons, and recognition is far more likely to form and endure.

Applied to AI, contestation asks for three elements working together. First, accessible challenge channels for consequential outcomes. People must know they can contest and how to do it: clear eligibility, low-friction filing, notice of rights in plain language, and support for affected parties. Second, a defined second look by a forum recognized as impartial. For private actors, that can be an internal appeal unit with structural separation and standards, paired with an external ombud, tribunal, or regulator for escalation; for public agencies, record-based administrative or judicial review. Third, remedies that actually fix things and teach institutions: reversals or corrections where warranted; reason letters that tie outcomes to enumerated factors; and, when appropriate, make-whole measures. Aggregated reporting then closes the loop by showing, in the open, how contestation changes outcomes and rules over time.

Design details matter for recognition. Timelines should be published and short; silence should not default to denial. Reason letters should reuse the vocabulary introduced in the rulebook, connect facts to listed considerations, and explain how competing values were balanced. Reviewers need guaranteed access to the information required for a meaningful second look, including model- or system-level materials when individual reasoning depends on them. There should be a duty to respond to substantial points and a public register of appeals, outcomes, and resulting rule amendments. These practices render voice, neutrality, respect, and explanation visible at the very points where decisions are felt.

Experience also warns against performativity. Organizations can accrue deference by showcasing procedures that change little on the ground—what Edelman calls legal endogeneity (Edelman, 2016). In the platform context, even prominent oversight bodies have improved transparency while revealing how dependence on authorizing firms can cabin independence, data access, and follow-through (Douek, 2024). Contestation contributes to legitimacy only if the second look is more than theatrical. Jurisdiction should not be at the provider's discretion; reviewers must have non-derogable access to the record; and there must be visible pathways from decisions to remedies and policy change. Absent these, appeals teach publics that "process" is a performance.

Public-sector automation offers both warnings and templates. Where automated adjudication lacked reasons and review, crises of recognition followed (Calo & Citron, 2021). Conversely, record-based second looks, explanation duties, and structured participation—documented across agency practice in the ACUS study—translate well to AI settings when adapted to the relevant venue (Engstrom, Ho, Sharkey, & Cuéllar, 2020). The lesson is consistent: contestation is not an add-on; it is the mechanism by which reasons bind.

Finally, contestation should feed back into rule formation. Petitions for amendment, periodic themed reviews of recurring issues, and public statements explaining how appeals have refined policies connect individual redress to systemic improvement. This is where the strands of the Article rejoin: integration locates authorship; familiarity provides the shared vocabulary and public reasons; contestation proves, case by case, that those reasons bind. The practical test is straightforward: can a person affected by an AI-backed decision see how to challenge it, obtain a prompt and impartial reconsideration, and receive an explanation and remedy that make sense in the publicly declared terms? If yes, contestation is doing its work.

## Conclusion

AI already governs, and it will govern more. If we treat technical performance as sufficient, we repeat a familiar error: systems can work and still be rejected because people can't see who authored the rules, why they bind, or how to get a credible second look. This Article's claim is straightforward: legitimacy is recognition—an audience-side judgment that power is exercised rightfully. Law can't create that recognition by fiat, but it can organize the conditions where recognition is likely. Seat rule-setting where the polity already locates authority; make reasons public in forms that fit local civic epistemologies; and guarantee contestation that's real rather than theatrical.

The analysis separated, then recombined, law's contributions. Thin legality supplies visible form: publicity, stability, prospectivity, consistency, professionalized routines. These signals matter; they make governance legible and dampen arbitrariness. Thick legality adds what form alone can't: public authorship and ownership of the governing rule-sets, audience-facing reasons, and a credible second look. Publicly authored AI rule-sets are one way to institutionalize thick legality for contemporary systems; they aren't the only way, but they show how authority can travel with the forum rather than the firm.

But none of this is automatic. These tools contribute to legitimacy where scope conditions are met: representation is credible, reasons are intelligible to ordinary audiences, and review has teeth. Otherwise, the same forms become ceremony—policy binders without constraint, "explanations" without reasons, appeals that never change outcomes. Legitimacy is local; the principles travel, the specific forms must be adapted. The point of offering principles—integration, familiarity, and contestation—is to guide that adaptation across jurisdictions and sectors without pretending one global design could secure recognition everywhere.

For policy, the three principles developed in Part IV supply a coherent program. Integration anchors AI rule-setting in venues a polity already recognizes as authoritative. Familiarity makes rules and reasons accessible in locally credible forms so people can anticipate outcomes and navigate governance. Contestation guarantees a credible second

look—voice, impartial review, and remedies that actually fix things. These principles are portable: jurisdictions can adapt them to local institutional arrangements and civic epistemologies while preserving the legitimacy logic they encode.

For theory, reframing AI governance around rightful authority does more than clarify what success looks like—it changes what regulatory tools we reach for and how we measure success. A performance-only frame treats resistance as ignorance or irrationality to be managed through better communication or iterative technical fixes. A legitimacy frame treats resistance as a signal that authorization is missing, reasons are opaque, or contestation is unavailable, and directs institutional design toward the conditions under which publics can form stable beliefs about rightful rule. The practical test is lived experience: Do people know where the rules are? Can they anticipate outcomes well enough to act? Do they know who to petition when rules should change? Can they challenge decisions and sometimes win? When those indicators are present, legitimacy is doing its work.

Without this, we face a predictable cycle: resistance, backlash, regulatory whiplash, and ultimately systems that perform yet cannot be governed because publics refuse to recognize their authority. If adopted, the principles offered here provide a path to govern AI in ways that publics recognize as rightful—avoiding a replay of legitimacy crises we already know too well, and making room for the technical work of alignment to matter in the places where people actually live with AI.